\documentclass[conference]{IEEEtran}
\IEEEoverridecommandlockouts
\usepackage{cite}
\usepackage{amsmath,amssymb,amsfonts}
\usepackage{algorithmic}
\usepackage{graphicx}
\usepackage{textcomp}

\usepackage{pgfplots}
\usepackage{pgfplotstable} % Для работы с CSV
\pgfplotsset{compat=1.18}
\usepackage{pgf}
\usetikzlibrary{pgfplots.fillbetween}
\usepackage{subcaption} % Для создания подфигур
\usepackage{url}
\usepackage{xfrac}

\usepackage{xcolor}
\usepackage[table,xcdraw]{xcolor}
\usepackage{tabularx}
\newcolumntype{Y}{>{\centering\arraybackslash}X}
\newcolumntype{P}[1]{>{\centering\arraybackslash}p{#1}}
\usepackage{subcaption} % Для subtable

\usepackage{seqsplit}

\usepackage{booktabs}

\def\BibTeX{{\rm B\kern-.05em{\sc i\kern-.025em b}\kern-.08em
    T\kern-.1667em\lower.7ex\hbox{E}\kern-.125emX}}
\begin{document}

\title{From Impermanent Loss to Sustainable Gain: Quantifying Profitability Zones for Liquidity Providers on DEX
% From Defaults to VaR: \\ Machine Learning in DeFi Lending Risk Analysis\\
}

\author{
\IEEEauthorblockN{
  Ignat Melnikov\IEEEauthorrefmark{1},
  Roman Vlasov\IEEEauthorrefmark{2},
  Vladimir Gorgadze\IEEEauthorrefmark{2}\IEEEauthorrefmark{3},
  Andrey Seoev\IEEEauthorrefmark{4}, %  andrei.seoev@mev-x.com
  Yury Yanovich\IEEEauthorrefmark{1}
}
% --- 1) Affiliation #1 ---
\IEEEauthorblockA{\IEEEauthorrefmark{1}%
  Skolkovo Institute of Science and Technology, Moscow, Russia \\
  % \textbf{Irina Lebedeva:} ORCID 0000-0001-5747-9284 \\
  % \textbf{Yury Yanovich:} ORCID 0000-0003-4651-7585 \\
  % \textbf{Ignat Melnikov:} ORCID 0009-0002-7804-8955 \\
  % \textbf{George Ovchinnikov:} g.ovchinnikov@skoltech.ru
}
\IEEEauthorblockA{\IEEEauthorrefmark{2}%
  Moscow Institute of Physics and Technology, Moscow, Russia \\
}
\IEEEauthorblockA{\IEEEauthorrefmark{3}%
  IDEAS: Inter-Disciplinary \& Advanced Studies Center, Moscow, Russia \\
}
\IEEEauthorblockA{\IEEEauthorrefmark{4}%
  MEV-X, Moscow, Russia 
}
% \vspace{-1cm}
}

\maketitle

\begin{abstract}
Decentralized Finance (DeFi) is a rapidly evolving segment of blockchain technology that enables a transformative approach to financial services through Web3 applications. By leveraging smart contracts, DeFi allows developers to build flexible and innovative financial instruments. Among the most prominent DeFi primitives by liquidity are decentralized exchange~(DEX) swap protocols~(such as Uniswap, Curve, and Balancer) that facilitate fast token-to-token exchanges. However, new exchange mechanisms also introduce new market inefficiencies that can be systematically exploited by arbitrageurs. This paper focuses on swap protocols based on the Automated Market Maker~(AMM), where the product of reserves is preserved as an invariant. We analyze the interaction between arbitrageurs and AMM liquidity pools and develop a mathematical model grounded in empirical pool configurations. Using this model, we derive bounds on the joint revenue of liquidity providers~(LPs) and arbitrageurs, propose a method to estimate the expected number of blocks until the occurrence of Impermanent Loss~(IL), and obtain a lower bound on the pool fee required to achieve a fixed target probability of staying in the Impermanent Gain (IG) zone within a block. The proposed framework extends existing LP risk-assessment methodologies by quantifying symbiotic profitability zones, providing a principled basis for fee selection that aligns LP-arbitrageur incentives and enhances market stability.
\end{abstract}

% 3...6 keywords
\begin{IEEEkeywords}
Arbitrage, Blockchain, Constant Product Market Maker, Decentralized Finance, Risk Assessment, Smart Contract, Impermanent Loss, Uniswap, Balancer
% Knowledge Discovery,  
% \TODO{need to check}
\end{IEEEkeywords}

\section{Introduction}

Decentralized Exchanges~(DEXs) powered by Automated Market Makers~(AMMs) are a foundational pillar of Decentralized Finance (DeFi)~\cite{Schar2020}, enabling trustless asset exchange without intermediaries~\cite{uniswapv2core2020, zetzsche2020decentralized, malamud2017decentralized,Hägele2024,Barbon2021}. Liquidity Providers~(LPs) are the essential counterparties in this system, depositing assets into liquidity pools in return for a share of trading fees. The market microstructure of these automated market makers has been analyzed from both adoption and efficiency perspectives~\cite{Capponi2021,Angeris2020}. However, a well-documented risk for LPs is \emph{Impermanent Loss} (IL)--the reduction in portfolio value compared to simply holding the assets, caused by adverse price movements relative to an external market~\cite{wang2022cyclic, hagele2024centralized}.

Conventional wisdom frames arbitrage--which aligns prices between different trading venues--as the primary driver of IL. Consequently, a significant line of research focuses on mitigating IL through mechanisms like dynamic fee adjustments~\cite{LebedevaDynFees, dynAMM, baggiani2025optimaldynamicfeesautomated}. However, this perspective is incomplete. Trading fees, the LP's reward, are generated predominantly by arbitrage activity. This creates a fundamental tension: the same trades that can cause IL also provide the fee income intended to compensate for it.

Recent work has introduced the concept of~\emph{Impermanent Gain}~(IG)--the scenario in which fees earned from an arbitrage trade exceed the associated IL, resulting in a net profit for the LP \cite{Vlasov2025Impact}. This occurs when the price deviation between the DEX and a reference market~(typically a CEX) is sufficiently small. This finding reframes the LP's risk-reward profile: rather than universally suffering from arbitrage, LPs can profit from it within a specific range of price discrepancies.

This paper moves from identifying the existence of IG to formally characterizing its~\emph{stable profitability zone} and demonstrating its practical feasibility. We develop a joint model of arbitrageur and LP profitability for Constant Function Market Makers~(CFMMs), including Uniswap V2 and Balancer. Within this model, we derive closed-form analytical bounds for Uniswap that define the exact price ratio interval where arbitrage is profitable for~\emph{both} the arbitrageur and the LP--a state of symbiotic equilibrium. Crucially, we validate this theoretical framework through a controlled on-chain experiment, deploying both baseline and private pools with a whitelisted arbitrageur, demonstrating the practical feasibility of collaborative systems where LPs can achieve stable returns with reduced impermanent loss.

Our core contributions are:
\begin{enumerate}
    \item \textbf{A Formal Model of Joint Profitability:} We formulate the arbitrageur's optimization problem for aligning DEX and CEX prices and solve it for Uniswap and Balancer, explicitly deriving the LP's resulting profit function.
    \item \textbf{Analytical Characterization of the IG Zone:} For Uniswap V2, we provide closed-form expressions for the boundaries of the price ratio region where LPs experience net impermanent gain.
    \item \textbf{Probabilistic Risk Framework:} By modeling external price dynamics as a Geometric Brownian Motion, we translate the IG zone boundaries into an upper bound for the one-block probability of IL, establishing a direct link between market volatility, protocol fees, and LP risk.
    \item \textbf{Fee Optimization Insight:} We demonstrate how this probabilistic bound can serve as a constraint for determining protocol fee parameters that align with a target LP risk tolerance.
    \item \textbf{Experimental Validation:} We conduct controlled on-chain experiments with private pools, demonstrating that 76\% of arbitrage transactions occur in win-win zones and revealing the regularization effect of fees against volatile arbitrage.
\end{enumerate}

Importantly, our experimental results reveal a nuanced trade-off: zero-fee pools generate 50\% more arbitrage activity but expose LPs to impermanent loss up to -20 MATIC, while even minimal fees (0.03\%) regularize arbitrage, stabilizing LP returns near zero IL. This suggests fee structures act as coordination mechanisms, not merely revenue sources.

By quantifying the stable zone of impermanent gain and demonstrating its practical viability, this work provides LPs with a clearer expectation of returns and offers protocol designers a mathematical tool for fee optimization based on measurable market risk. The subsequent sections detail our methodology (Section~\ref{sec:methods}), present the theoretical analysis (Section~\ref{sec:theoretical}), show experimental validation (Section~\ref{sec:exps}), and discuss the implications of our findings (Section~\ref{sec:conclusion}).

\section{Related Work}\label{sec:related_work}

Automated Market Makers (AMMs) enable decentralized token exchange by replacing the traditional order book with a liquidity pool and a pricing function implemented in a smart contract~\cite{zetzsche2020decentralized,uniswapv2core2020}. The quality and efficiency of these decentralized markets compared to centralized exchanges have been empirically studied~\cite{Barbon2021}, revealing trade-offs in market depth and price discovery. In a two-asset AMM, the pool holds reserves \((x,y)\) and defines a trading rule via an invariant, e.g., a curve \(F(x,y)=\text{const}\). Traders swap against the pool by moving the state along this curve: adding one asset requires removing the other so that the invariant is (approximately) preserved. The instantaneous exchange rate is given by the pool’s marginal price, which corresponds to the local slope of the invariant, \(p = -\,\mathrm{d}y/\mathrm{d}x\). Because this slope changes with the reserves, AMMs naturally exhibit slippage (larger trades move the price more). Liquidity providers (LPs) supply both assets to the pool and receive LP shares, earning a fraction of swap fees, but their returns depend on how prices evolve, since changes in the reserve ratio can lead to effects such as impermanent loss.

Impermanent Loss is classically defined as the difference between the value of a liquidity provider's position in an AMM and the value of holding the same assets outside the pool~\cite{Xu_2023, uniswapv2core2020, IL, Aigner2021, Loesch2021}. Mathematical formalizations have further refined this concept through the lens of "Greeks" for liquidity providers~\cite{Bardoscia2023} and generalized treatments across constant function market makers~\cite{Tangri2023}. When the relative market price of the two tokens changes, arbitrage trades rebalance the pool reserves to the new price, so the LP ends up with a different asset mix. This mix can be worth less than the original hold portfolio because of the AMM’s convex rebalancing rule. The loss is called \emph{impermanent} because it depends on the current price ratio and can shrink if prices revert, while accumulated swap fees may partially or fully offset it~\cite{Vlasov2025Impact}. This view, however, primarily captures the LP's perspective relative to a passive "hold" baseline. Recent research has critically expanded this understanding by introducing the symmetrical concept of \textbf{impermanent gain}, reframing IL not merely as a loss but as a transfer of value between counterparties in a trade. As shown by Kim et al.~\cite{Kim2022}, impermanent gain is the positive financial outcome for the AMM (and thus collectively for its LPs) that arises from the exact same price movement that causes an impermanent loss for an individual LP when evaluated against the hold strategy. This duality is most clearly articulated by Labadie, who demonstrates that for AMMs with a constant-product formula, the trader's slippage is mathematically equivalent to the AMM's negative IL--that is, the trader's loss from slippage is precisely the AMM's (and LPs') impermanent gain~\cite{Labadie2022}. This establishes that IL and slippage are two sides of the same coin, representing a fundamental wealth transfer from traders to the liquidity pool during non-infinitesimal swaps. Building on this, Lee and Kim~\cite{Lee2024b} provide a significant theoretical advancement by defining a sufficient condition for guaranteed impermanent gain, which they term \textbf{"origin crossing"}. They prove that a swap transaction results in a net impermanent gain for the liquidity pool if the trade moves the pool's reserve state \((x, y)\) in such a way that the post-trade state is on the opposite side of a line connecting the pre-trade state to the origin (i.e., the ray from the origin to the state changes quadrant). This provides a concrete, testable criterion for when LPs benefit from price divergence, moving the discussion from probabilistic fee offsets to deterministic transaction-based outcomes.

Popular AMM designs primarily differ in their choice of invariant, which directly shapes the impermanent loss (IL) profile. For instance, Uniswap V2 employs a constant product invariant~\cite{uniswapv2core2020}:
\begin{equation*}
F_u(x,y,K) = xy - K^2,
\end{equation*}
where $K$ is a constant determined by the initial pool reserves and maintained (up to fees) during trading. In contrast, Balancer extends this framework to a weighted constant product invariant~\cite{martinelli2019balancer}.
\begin{equation*}
    F_b(x,y,K) = x^{w_x}y^{w_y} - K^{w_x + w_y}, \space{} w_x+w_y = 1.
\end{equation*}
Comparative analyses like Kim et al.~\cite{Kim2024} show how different invariant curves (constant product, constant sum, stable swap) lead to distinct IL landscapes under the same market movements, informing LP choice of platform. Beyond basic invariant designs, concentrated liquidity markets such as Uniswap V3 have introduced more complex risk-return profiles for LPs. Analyses of these markets include predictable losses in concentrated liquidity~\cite{Cartea2023}, risks and returns of Uniswap V3 liquidity providers~\cite{Heimbach2022}, strategic liquidity provision in Uniswap V3~\cite{Fan2024}, mathematical derivations of concentrated liquidity mechanics~\cite{Richardson2024}, and thorough mathematical modeling of Uniswap v3 dynamics~\cite{Rigneault2023}.

A primary line of defense against net IL is the accumulation of swap fees. In~\cite{Vlasov2025Impact}, the authors study several AMM designs (UniswapV2, Balancer, and Curve). They emphasize that swap fees, when reinvested into the liquidity pool, make the pool invariant effectively non-constant over time. As a consequence, there exists a non-trivial region of impermanent gain for liquidity providers when the trade size is sufficiently small. For UniswapV2, the paper derives an exact expression for the threshold swap size, where \(\phi_1\) and \(\phi_2\) are the fees for the input and output tokens and \(\gamma_i = 1 - \phi_i\):
\begin{equation}\label{amax_uniswap}
    \Delta x^u_\text{max} = \frac{1-\gamma_1\gamma_2}{\gamma_1\,(2\gamma_2-1)} x
\end{equation}
It also provides an approximate bound that applies to all considered protocols, where \(p(x_0, K_0)\) denotes the pool’s marginal price at the initial moment of time:
\begin{equation}\label{amax_general}
    \Delta x_{\max} \approx 2(\phi_1+\phi_2)\,\frac{p(x_0,K_0)}{\left|\dfrac{\partial p(x_0,K_0)}{\partial x}\right|}
\end{equation}

The strategic setting of these fees is itself a complex optimization problem. Recent work explores \textbf{dynamic fee mechanisms} as a sophisticated tool for IL mitigation. Baggiani et al.~\cite{baggiani2025optimaldynamicfeesautomated} formulate the problem of determining optimal dynamic fees in a constant function market maker as a stochastic control problem. Their analysis reveals two distinct optimal regimes: a high-fee regime to deter arbitrage and preserve LP value, and a low-fee regime designed to attract beneficial "noise" trading volume and increase fee income, with the optimal fee adapting linearly to the pool's inventory and external price changes. This aligns with empirical observations in stable-swap markets, where fee compression is driven by competition, yet ultra-low fees can paradoxically attract more toxic, MEV-driven order flow~\cite{khailuk2025stateofstableswaps}. The broader ecosystem impact of MEV includes consensus instability and frontrunning in decentralized exchanges~\cite{Daian2020}, while systematic approaches to arbitrage in DEXs have been formalized~\cite{Boonpeam2021}. Game-theoretic models, such as those by Fritsch~\cite{Fritsch_2021}, further analyze fee competition between pools, proving the existence of equilibria where traders endogenously split order flow. Beyond fees, advanced strategies focus on \textbf{optimizing liquidity deployment}. For concentrated liquidity AMMs like Uniswap V3, Zeller et al. (2025) formalize the problem of selecting optimal liquidity provision intervals as a tractable stochastic optimization problem, balancing expected fee rewards against divergence loss and reallocation costs~\cite{IL2}. This approach aims to move LPs from heuristic methods to mathematically grounded strategies for profitable liquidity concentration.

Finally, the management of IL sits within the broader DeFi landscape where protocols expose sensitive hyperparameters. System outcomes are highly dependent on these choices because user behavior adapts endogenously. Thus, tuning parameters becomes a coupled optimization problem under strategic interaction rather than a simple offline calibration task. This principle is evident not only in AMM fee games~\cite{Fritsch_2021, dynAMM} but also in lending protocols, where parameters like collateral factors and interest rates are optimized through the lens of default probability and risk assessment~\cite{lending1, lending2}. 

The design of optimal liquidity provision mechanisms has been formalized through Myersonian frameworks that consider incentive compatibility and social welfare~\cite{Milionis2024}. However, liquidity provision in decentralized exchanges exhibits paradoxes where more providers can sometimes mean less effective liquidity due to fragmentation and coordination challenges~\cite{Capponi2023}. Broader DeFi risk measurement frameworks address both protocol-specific risks and systemic vulnerabilities~\cite{Bertomeu2024, Carter2021}.

In summary, the related work reveals impermanent loss as a central, multifaceted challenge in DeFi. The discourse has evolved from viewing it as a simple passive holding penalty to understanding it as one side of a value transfer (impermanent gain), with deterministic conditions for its occurrence. Mitigation strategies are advancing from static fee models to dynamic, optimized approaches for both fee setting and liquidity placement, all within a complex, game-theoretic ecosystem of interacting agents.

\section{Methodology}\label{sec:methods}

Our methodology aims to establish a quantitative framework for analyzing the symbiotic relationship between liquidity providers (LPs) and arbitrageurs in AMM-based DEXs. We conceptualize three distinct price ratio zones that determine profitability outcomes for both parties, as illustrated in Figure~\ref{fig:profits}. By combining this zonal analysis with stochastic modeling of price movements, we derive probabilistic measures of IL risk and corresponding fee optimization strategies.

\subsection{Joint Profitability Zones}

The core of our approach lies in identifying three mutually exclusive regions defined by the ratio between the external market price \(p_{\text{cex}}\) and the DEX pool's marginal price \(p_{\text{dex}}\):

\begin{enumerate}
    \item \textbf{No-Arbitrage Zone:} When \(|p_{\text{cex}}/p_{\text{dex}} - 1|\) is very small, arbitrage opportunities are insufficient to cover transaction costs. In this zone, no arbitrage transactions occur, and LPs neither gain nor lose from price movements.
    
    \item \textbf{Impermanent Gain (IG) Zone:} For moderate price discrepancies beyond the no-arbitrage threshold but within a specific bound, arbitrage becomes profitable. Crucially, the arbitrage trade volume in this region is small enough that the fees collected by LPs exceed the IL caused by the price realignment. This creates a \emph{win-win} scenario where both arbitrageurs and LPs profit.
    
    \item \textbf{Impermanent Loss (IL) Zone:} When price discrepancies are large, arbitrage remains profitable for arbitrageurs, but the required trade volume is substantial. In this zone, the resulting IL outweighs the fee income for LPs, leading to a net loss.
\end{enumerate}

The boundaries between these zones depend on the pool's fee structure (\(\phi_1, \phi_2\)) and invariant curve parameters. For the constant product AMMs we analyze (Uniswap V2 and Balancer), these boundaries are symmetric around 1 when expressed as price ratios. Specifically, if we denote the upper boundary as \(\tau > 1\), then the lower boundary is \(1/\tau\), creating an interval \([1/\tau, \tau]\) where arbitrage results in impermanent gain for LPs. This symmetry arises from the mathematical structure of these AMMs and is not a general requirement for all market makers.

\subsection{Stochastic Risk Assessment}

To translate zonal boundaries into practical risk metrics, we model the evolution of the external price \(p_{\text{cex}}\) as a Geometric Brownian Motion (GBM). Given the IG zone boundaries \([1/\tau, \tau]\), we compute the probability that a price movement within one block time \(\Delta t\) crosses from the IG zone into the IL zone. This one-block IL probability, \(P_{\text{IL}}\), serves as a key risk metric for LPs.

Under the GBM assumption, the distribution of price ratios is log-normal, allowing us to derive an upper bound for \(P_{\text{IL}}\). This bound depends explicitly on market volatility \(\sigma\), block time \(\Delta t\), and the fee-dependent boundary parameter \(\tau\).

\subsection{Fee Parameter Optimization}

The final component of our methodology links the risk metric to protocol design. Given a target maximum acceptable IL probability \(\xi\) (e.g., 1\% per block), we solve for the minimum fee parameters \((\phi_1, \phi_2)\) that satisfy \(P_{\text{IL}} \leq \xi\). This provides a principled approach to fee setting that aligns protocol parameters with LP risk tolerance, moving beyond heuristic or purely competitive fee selection.

\section{Theoretical Analysis}\label{sec:theoretical}

In this section, we present the detailed mathematical derivations underlying our methodology. We begin with the arbitrageur's optimization problem, derive explicit profitability zones for Uniswap V2 and Balancer, and establish the probabilistic risk framework.

\subsection{Arbitrageur Model}

An arbitrageur maximizes profit by exploiting price differences between a DEX pool and an external CEX. We assume the CEX is infinitely liquid at price \(p_{\text{cex}}\), where \(p_{\text{cex}}\) denotes the external market price of the asset pair. The DEX follows a constant function market maker curve with fees, and its marginal price is denoted by \(p_{\text{dex}}\), representing the instantaneous exchange rate implied by the pool reserves. Let \(\gamma_i = 1 - \phi_i\) where \(\phi_i\) is the fee rate for token \(i\), so that \(\gamma_i\) represents the effective fraction of input remaining after fees. The arbitrageur's profit when \(p_{\text{dex}} < p_{\text{cex}}\) (buy on DEX, sell on CEX) is:

\begin{equation}\label{arb_problem}
    \Pi_{\text{arb}} = -dx + p_{\text{cex}} \cdot dy(dx),
\end{equation}
where \(dy(dx)\) is the output from the DEX for input \(dx\), accounting for fees. The arbitrageur chooses \(dx\) to maximize \(\Pi_{\text{arb}}\).

\subsubsection{Balancer}

For Balancer with weights \(w_x, w_y\) (\(w_x + w_y = 1\)), the trading function with fees is:

\begin{equation*}
    (x + \gamma_1 dx)^{w_x} (y - dy)^{w_y} = x^{w_x} y^{w_y}.
\end{equation*}

Solving for the optimal arbitrage amounts yields:

\begin{equation}\label{balancer_arb}
\begin{cases}
    dx^{*} = \frac{x}{\gamma_1}\left[\left(\frac{\gamma_1\gamma_2\,p_{\text{cex}}}{p_{\text{dex}}}\right)^{w_y}- 1\right], & p_{\text{cex}} > \frac{p_{\text{dex}}}{\gamma_1\gamma_2} \\
    dy^{*} = \frac{y}{\gamma_1}\left[\left(\frac{\gamma_1\gamma_2\,p_{\text{dex}}}{p_{\text{cex}}}\right)^{w_x}- 1\right], & p_{\text{cex}} < \gamma_1\gamma_2 p_{\text{dex}}
\end{cases}.
\end{equation}

While Balancer generalizes the constant product invariant through asset weights, this added flexibility reduces analytical tractability. In particular, closed-form expressions for the joint profitability (IG) zone boundaries are generally not available due to the nonlinear dependence on the weight parameters. As a result, in this work we focus on deriving explicit analytical results for Uniswap V2, while the corresponding results for Balancer are obtained numerically based on the arbitrage optimization framework.

\subsubsection{Uniswap V2}

Uniswap V2 is a special case of Balancer with \(w_x = w_y = 0.5\). Substituting into (\ref{balancer_arb}) gives:

\begin{equation}\label{uniswap_arb}
\begin{cases}
    dx^{*} = \frac{x}{\gamma_1}\left(\sqrt{\frac{\gamma_1\gamma_2\,p_{\text{cex}}}{p_{\text{dex}}}}-1\right), & p_{\text{cex}} > \frac{p_{\text{dex}}}{\gamma_1\gamma_2} \\
    dy^{*} = \frac{y}{\gamma_1}\left(\sqrt{\frac{\gamma_1\gamma_2\,p_{\text{dex}}}{p_{\text{cex}}}}-1\right), & p_{\text{cex}} < \gamma_1\gamma_2 p_{\text{dex}}
\end{cases}.
\end{equation}

\subsection{Joint Profitability Zones}

The LP experiences impermanent gain when the arbitrage volume is below a threshold \(\Delta x_{\max}\) (or \(\Delta y_{\max}\)). For Uniswap V2, this threshold is given by \cite{Vlasov2025Impact}:

\begin{equation}\label{lp_borders}
    \Delta x^\text{LP} = x\frac{1-\gamma_1\gamma_2}{\gamma_1\,(2\gamma_2-1)}, \quad 
    \Delta y^\text{LP} = y\frac{1-\gamma_1\gamma_2}{\gamma_1\,(2\gamma_2-1)}.
\end{equation}

The joint profitability zone occurs when the arbitrageur's optimal trade (\ref{uniswap_arb}) is less than the LP's gain threshold (\ref{lp_borders}). Solving these inequalities yields the price ratio interval for win-win outcomes:

\begin{equation}\label{joint_zone}
    \frac{p_{\text{cex}}}{p_{\text{dex}}} \in \left[\frac{\gamma_1(2\gamma_2-1)^2}{\gamma_2(2-\gamma_1)^2}, \frac{\gamma_2(2-\gamma_1)^2}{\gamma_1(2\gamma_2-1)^2}\right] = [1/\tau, \tau],
\end{equation}
where \(\tau = \frac{\gamma_2(2-\gamma_1)^2}{\gamma_1(2\gamma_2-1)^2}\). Note that the interval is symmetric around 1 in logarithmic space (\(\log(1/\tau) = -\log(\tau)\)), which is a specific property of constant product AMMs with the fee structure considered. This symmetry simplifies the probabilistic analysis but is not a general requirement for all AMM designs.

\subsection{Probability of Impermanent Loss}

Assuming GBM dynamics for \(p_{\text{cex}}\) with drift \(\mu\) and volatility \(\sigma\), the one-block transition probability density is log-normal. The probability of exiting the IG zone in one block, starting from price ratio \(y = p_{\text{cex}}/p_{\text{dex}}\), is:

\begin{multline}\label{pil_exact}
    P_{\text{IL}}(y) = 1 - \Phi\left(\frac{\ln(1/\tau) - (\ln y + (\mu-\tfrac12\sigma^2)\Delta t)}{\sigma\sqrt{\Delta t}}\right) \\
    + \Phi\left(\frac{\ln \tau - (\ln y + (\mu-\tfrac12\sigma^2)\Delta t)}{\sigma\sqrt{\Delta t}}\right),
\end{multline}
where \(\Phi\) is the standard normal CDF.

An upper bound for \(P_{\text{IL}}\) is obtained by evaluating at the worst-case starting point within the IG zone. For \(\mu = 0\), this occurs at the boundaries:

\begin{multline}\label{pil_bound}
    P_{\text{IL}}^{\text{up}}(\gamma_1, \gamma_2) = 1 - \Phi\left(\frac{\ln\left(\frac{(2-\gamma_1)^2}{\gamma_1^{2}(2\gamma_2-1)^2}\right) + \tfrac12\sigma^2\Delta t}{\sigma\sqrt{\Delta t}}\right) \\
    + \Phi\left(\frac{\ln\left(\frac{(2\gamma_2-1)^2}{\gamma_2^{2}(2-\gamma_1)^2}\right) + \tfrac12\sigma^2\Delta t}{\sigma\sqrt{\Delta t}}\right).
\end{multline}

\subsection{Fee Optimization}

Given a target IL probability \(\xi\), we solve for the fee parameters that satisfy \(P_{\text{IL}}^{\text{up}} \leq \xi\). For one-sided fees (\(\phi_1 = \phi\), \(\phi_2 = 0\)), this reduces to a one-dimensional root-finding problem. The solution provides the minimum fee required to achieve the desired risk level given market volatility \(\sigma\).

\section{Experimental Results}\label{sec:exps}

We conduct two sets of experiments: (1) simulation studies to validate our theoretical framework and illustrate the methodology, and (2) real-world on-chain experiments with private pools to demonstrate practical feasibility.

\subsection{Illustration of Methodology}

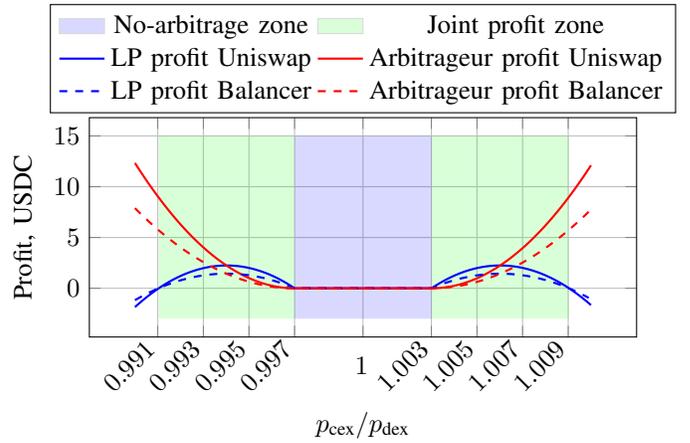
\begin{figure}[t]
\centering
\begin{tikzpicture}
\begin{axis}[
  width=\linewidth,
  height=4.5cm,
  grid=both,
  xlabel={\({p_\text{cex}}/{p_\text{dex}}\)},
  ylabel={Profit, USDC},
  legend style={at={(0.5,1.02)}, anchor=south, legend columns=2},
  cycle list={}, % ✅ FIXED: Added missing comma here
  xtick={0.991,0.993,0.995,0.997,1,1.003,1.005,1.007,1.009},
  x tick label style={anchor=east,  yshift=-11pt, xshift=3pt},
  xticklabel={
    \pgfmathparse{abs(\tick-1) < 0.0005}%
    \ifnum\pgfmathresult=1
      \makebox[0pt][c]{\pgfmathprintnumber[fixed,precision=3]{\tick}}%
    \else
      \rotatebox{45}{\pgfmathprintnumber[fixed,precision=3]{\tick}}%
    \fi
  },
]

% ---- legend (fixed mapping) ----
\addlegendimage{area legend, fill=blue, draw=none, opacity=0.15}
\addlegendentry{No-arbitrage zone}

\addlegendimage{area legend, fill=green, draw=none, opacity=0.15}
\addlegendentry{Joint profit zone}

\addlegendimage{no marks, thick, solid, blue}
\addlegendentry{LP profit Uniswap}

\addlegendimage{no marks, thick, solid, red}
\addlegendentry{Arbitrageur profit Uniswap}

\addlegendimage{no marks, thick, dashed, blue}
\addlegendentry{LP profit Balancer}

\addlegendimage{no marks, thick, dashed, red}
\addlegendentry{Arbitrageur profit Balancer}

% ---- shaded zones (no legend entries here) ----
\addplot [name path=topNP, draw=none, domain=0.997:1.0030090270812437] {15};
\addplot [name path=botNP, draw=none, domain=0.997:1.0030090270812437] {-3};
\addplot [fill=blue, draw=none, opacity=0.15, forget plot]
  fill between[of=topNP and botNP];

\addplot [name path=topG1, draw=none, domain=1.0030090270812437:1.009036108324975] {15};
\addplot [name path=botG1, draw=none, domain=1.0030090270812437:1.009036108324975] {-3};
\addplot [fill=green, draw=none, opacity=0.15, forget plot]
  fill between[of=topG1 and botG1];

\addplot [name path=topJP, draw=none, domain=0.9910448117263363:0.997] {15};
\addplot [name path=botJP, draw=none, domain=0.9910448117263363:0.997] {-3};
\addplot [fill=green, draw=none, opacity=0.15, forget plot]
  fill between[of=topJP and botJP];

% ---- curves (no legend entries here) ----
\addplot+[no marks, thick, solid, forget plot, blue] table[
  col sep=comma,
  x=price,
  y=lp_profit
]{data/uniswap.csv};

\addplot+[no marks, thick, solid, forget plot, red] table[
  col sep=comma,
  x=price,
  y=arb_profit
]{data/uniswap.csv};

\addplot+[no marks, thick, dashed, forget plot, blue] table[
  col sep=comma,
  x=price,
  y=lp_profit
]{data/balancer_08.csv};

\addplot+[no marks, thick, dashed, forget plot, red] table[
  col sep=comma,
  x=price,
  y=arb_profit
]{data/balancer_08.csv};

\end{axis}
\end{tikzpicture}
\caption{LP and arbitrageur profit functions for Uniswap (solid) and Balancer with \(w_x=0.2\), \(w_y=0.8\) (dashed). The blue shaded area indicates the no-arbitrage zone, while green areas represent the joint profitability (IG) zone.}
\label{fig:profits}
\end{figure}

We first apply our framework to simulated Uniswap V2 and Balancer pools with parameters matching real volatile pairs (Table~\ref{tab:rad-usdc}). Figure~\ref{fig:profits} visualizes the three profitability zones and the profit functions for both LPs and arbitrageurs. The green region represents the win-win zone where both parties profit, validating our theoretical bounds.

\begin{table}[t]
\centering
\caption{Simulation parameters based on USDC/RAD pool}
\label{tab:rad-usdc}
\begin{tabularx}{\linewidth}{|P{1.5cm}|Y|Y|Y|Y|Y|}
\hline
\rowcolor[HTML]{EFEFEF}
\textbf{Pool} &
\multicolumn{5}{c|}{\cellcolor[HTML]{EFEFEF}\textbf{Ethereum Address}} \\ \hline

USDC/RAD &
\multicolumn{5}{c|}{\url{0x8C1c499b1796D7F3C2521AC37186B52De024e58c}} \\ \hline

\rowcolor[HTML]{EFEFEF}
\textbf{\(\sigma\)} &
\textbf{\(\mu\)} &
\textbf{\(\gamma_1\)} &
\textbf{\(\gamma_2\)} &
\textbf{\(x\)} &
\textbf{\(y\)} \\ \hline

0.0027 & 0 & 0.997 & 1 & 997\,348 & 3\,751\,882 \\ \hline
\end{tabularx}
\end{table}

\subsubsection{IL Block Distribution}

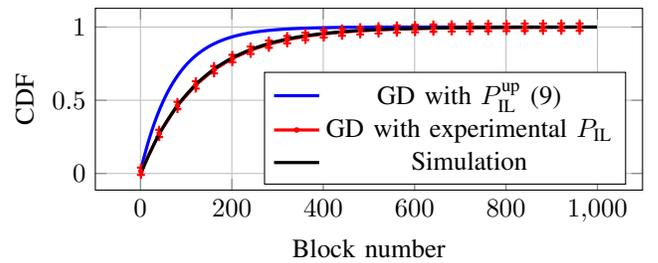
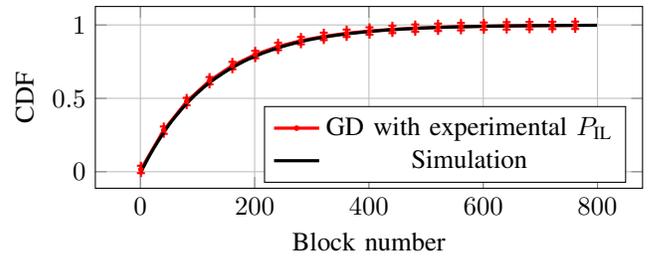
\begin{figure}
\begin{subfigure}[t]{\linewidth}
\centering
\begin{tikzpicture}
\begin{axis}[
  width=\linewidth,
  height=4cm,
  grid=both,
  xlabel={Block number},
  ylabel={CDF},
  legend pos=south east,
  cycle list={}
]

% theory
\addplot+[no marks, very thick, solid, blue]
table[col sep=comma, x=x, y=theory]{data/cdf_uniswap.csv};
\addlegendentry{GD with \(P_\text{IL}^\text{up}\)~\eqref{pil_bound}}

% exp_prob with crosses + error bars
\addplot+[
  mark=x, mark size=1pt, very thick, red, each nth point=40,
  error bars/.cd, y dir=both, y fixed=0.025, error mark=+, error mark options={mark size=1.5pt, thick}
]
table[col sep=comma, x=x, y=exp_prob]{data/cdf_uniswap.csv};
\addlegendentry{GD with experimental \(P_\text{IL}\)}

% sim
\addplot+[no marks, very thick, solid, black]
table[col sep=comma, x=x, y=sim]{data/cdf_uniswap.csv};
\addlegendentry{Simulation}

\end{axis}
\end{tikzpicture}
\caption{Uniswap V2.}
\end{subfigure}

\vspace{0.6em}

\begin{subfigure}[t]{\linewidth}
\centering
\begin{tikzpicture}
\begin{axis}[
  width=\linewidth,
  height=4cm,
  grid=both,
  xlabel={Block number},
  ylabel={CDF},
  legend pos=south east,
  cycle list={}
]

% exp_prob with crosses + error bars
\addplot+[
  mark=x, mark size=1pt, very thick, red, each nth point=40,
  error bars/.cd, y dir=both, y fixed=0.025, error mark=+, error mark options={mark size=1.5pt, thick}
]
table[col sep=comma, x=x, y=exp_prob]{data/cdf_balancer.csv};
\addlegendentry{GD with experimental \(P_\text{IL}\)}

% sim
\addplot+[no marks, very thick, solid, black]
table[col sep=comma, x=x, y=sim]{data/cdf_balancer.csv};
\addlegendentry{Simulation}

\end{axis}
\end{tikzpicture}
\caption{Balancer, \(w_x=0.2\), \(w_y=0.8\).}
\end{subfigure}
\caption{CDF of the block number until first IL occurrence. The geometric distribution (GD) model with experimental \(P_{\text{IL}}\) closely matches simulations.}
\label{cdfs}
\end{figure}

We simulate price paths under GBM to estimate the distribution of blocks until first IL occurrence. Specifically, we simulate $10,000$ independent price trajectories following a Geometric Brownian Motion with drift \(\mu\) and volatility \(\sigma\), using a discrete-time approximation with block interval \(\Delta t = 12\) seconds. For each trajectory, we record the number of blocks until the first transition from the IG zone to the IL zone occurs. Figure~\ref{cdfs} compares the empirical cumulative distribution function (CDF) with theoretical predictions. The geometric distribution model with experimentally estimated \(P_{\text{IL}}\) closely matches simulation results, while the conservative upper bound \(P_{\text{IL}}^{\text{up}}\) provides a safe overestimate.

\subsubsection{Optimal Fee Analysis}

\begin{figure}[t]
\centering
\begin{tikzpicture}
\begin{axis}[
  width=\linewidth,
  height=4.5cm,
  grid=both,
  xlabel={Standard Deviation per block},
  ylabel={Fee},
legend style={at={(0.58,1.02)}, anchor=south, legend columns=3},
  cycle list={}
]

\addlegendimage{thick, solid, blue}
\addlegendentry{Balancer}

\addlegendimage{thick, solid, red}
\addlegendentry{Uniswap}

\addlegendimage{thick, solid, black}
\addlegendentry{Theory}

\addlegendimage{thick, dashdotted, black}
\addlegendentry{\(\xi = 0.001\)}

\addlegendimage{thick, dashed, black}
\addlegendentry{\(\xi = 0.01\)}

\addlegendimage{thick, dotted, black}
\addlegendentry{\(\xi = 0.05\)}

\addplot+[
  mark=x, mark size=1pt, very thick, blue, dashdotted,
  error bars/.cd, y dir=both, y fixed=0.0001, error mark=+, error mark options={mark size=2.5pt, very thick}
]
table[col sep=comma, x=std, y=fees_balancer]{data/0.001_sig_fees.csv};
\addplot+[
  mark=x, mark size=1pt, very thick, red, dashdotted,
  error bars/.cd, y dir=both, y fixed=0.0001, error mark=+, error mark options={mark size=1.5pt, thick}
]
table[col sep=comma, x=std, y=fees_uniswap]{data/0.001_sig_fees.csv};
\addplot+[no marks, very thick, solid, black, dashdotted]
table[col sep=comma, x=std, y=fees_th]{data/0.001_sig_fees.csv};

\addplot+[
  mark=x, mark size=1pt, very thick, dashed, blue,
  error bars/.cd, y dir=both, y fixed=0.0001, error mark=+, error mark options={mark size=2.5pt, very thick}
]
table[col sep=comma, x=std, y=fees_balancer]{data/0.01_sig_fees.csv};
\addplot+[
  mark=x, mark size=1pt, very thick, dashed, red,
  error bars/.cd, y dir=both, y fixed=0.0001, error mark=+, error mark options={mark size=1.5pt, thick}
]
table[col sep=comma, x=std, y=fees_uniswap]{data/0.01_sig_fees.csv};
\addplot+[no marks, very thick, dashed, black]
table[col sep=comma, x=std, y=fees_th]{data/0.01_sig_fees.csv};

\addplot+[
  mark=x, mark size=1pt, very thick, dotted, blue,
  error bars/.cd, y dir=both, y fixed=0.0001, error mark=+, error mark options={mark size=2.5pt, very thick}
]
table[col sep=comma, x=std, y=fees_balancer]{data/0.05_sig_fees.csv};
\addplot+[
  mark=x, mark size=1pt, very thick, dotted, red,
  error bars/.cd, y dir=both, y fixed=0.0001, error mark=+, error mark options={mark size=1.5pt, thick}
]
table[col sep=comma, x=std, y=fees_uniswap]{data/0.05_sig_fees.csv};
\addplot+[no marks, very thick, dotted, black]
table[col sep=comma, x=std, y=fees_th]{data/0.05_sig_fees.csv};

\end{axis}
\end{tikzpicture}
\caption{Minimum fee required to achieve target IL probability \(\xi\) at different volatility levels. Theoretical bounds (black lines) provide close approximations to simulation results (points with error bars).}
\label{fig:fees}
\end{figure}
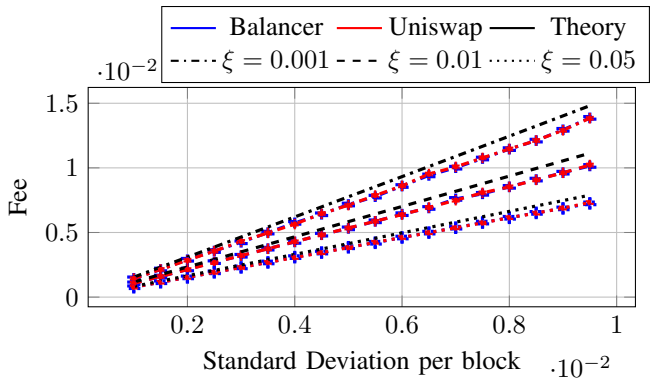

Figure~\ref{fig:fees} shows the minimum fee required to achieve target IL probabilities \(\xi = \{0.001, 0.01, 0.05\}\) across different volatility levels. Our theoretical bound provides a practical approximation for protocol designers, with close alignment to experimental results.

\subsection{On-Chain Experiment with Private Pools}

To demonstrate the practical feasibility of our theoretical framework, we conducted a controlled on-chain experiment using two Algebra V3 (Uniswap V3-style) concentrated liquidity pools deployed on the Polygon network~\cite{Kanani2018}. The experiment ran from December 1st to December 26th, 2025, providing sufficient data to analyze the dynamics of joint profitability under real market conditions.

\subsubsection{Experimental Setup}

We deployed two pools with identical parameters except for their fee structure. Both pools used the USDT/WMATIC pair, where WMATIC represents the wrapped version of Polygon's native MATIC token (wrapping is standard practice in DEXs prior to Uniswap V4 for technical simplicity and uniform token interfaces). Pool A implemented a 0.03\% swap fee, while Pool B served as a zero-fee alternative. The small fee in Pool A was intentionally chosen to create a narrower IG zone, making impermanent loss more probable--thereby testing our risk framework under challenging conditions--and to test the regularization effect of fees against low-profit, volatile arbitrage. This low-fee approach also aligns with recent trends in stable-swap markets, where fee compression has become prevalent due to intense competition \cite{Khailuk2025}.

To establish exclusive arbitrage access--a critical requirement for testing our collaborative LP-arbitrageur model--we implemented a whitelist mechanism that restricted swap permissions to our designated arbitrage bot. This setup ensured that all price alignment transactions originated from our controlled arbitrage strategy, eliminating interference from external arbitrageurs and MEV bots. This design choice is methodologically motivated. In real-world DEX environments, arbitrage is highly competitive and often dominated by MEV-driven strategies, as documented in prior work~\cite{Daian2020, Boonpeam2021, Fritsch_2021, Capponi2023}. By restricting access to a single arbitrageur, we intentionally remove competitive and MEV-related effects in order to isolate the core LP--arbitrageur interaction. This allows us to evaluate the underlying mechanism in a controlled setting.

Both pools were configured with liquidity concentrated in a single wide price range [0.0001, 10000] to approximate the behavior of a Uniswap V2 pool while maintaining the gas efficiency benefits of concentrated liquidity. We deployed \$3,000 worth of liquidity in each pool, split appropriately between USDT and WMATIC to establish an initial price aligned with the external market. Table~\ref{tab:private_pools} summarizes the key experimental parameters.

\begin{table}[t]
\centering
\caption{Private pool experiment parameters}
\label{tab:private_pools}
\begin{tabularx}{\linewidth}{|l|Y|Y|}
\hline
\rowcolor[HTML]{EFEFEF} \textbf{Parameter} & \textbf{Pool A (0.03\% fee)} & \textbf{Pool B (0\% fee)} \\ \hline
Address & \seqsplit{0x1C890425Cc2FFF65891Be6c87A07E9Dd4613C658} & \seqsplit{0x5A5E71Dc833e49F4cE29F72518d0D4877aA581bB} \\ \hline
Token Pair & USDT/WMATIC & USDT/WMATIC \\ \hline
Liquidity & \$3,000 & \$3,000 \\ \hline
Fee Tier & 0.03\% & 0\% \\ \hline
Range & [0.0001, 10000] & [0.0001, 10000] \\ \hline
Duration & Dec 1--26, 2025 & Dec 1--26, 2025 \\ \hline
Whitelisted & Arbitrage bot only & Arbitrage bot only \\ \hline
\end{tabularx}
\end{table}

Our arbitrage bot continuously monitored both pools and external DEX pools. When price discrepancies exceeded the no-arbitrage threshold, the bot executed optimal arbitrage trades according to the strategy derived in Section~\ref{sec:theoretical}. All transactions, profits, and impermanent loss events were tracked on-chain for subsequent analysis.

\subsubsection{Results and Analysis}

\begin{figure}[t]
    \centering
    \begin{tikzpicture}
    \begin{axis}[
        width=\linewidth,
        height=4cm,
        ylabel={Count},
        ymin=0,
        enlarge x limits=0.25,
        ymode=log,
        xtick=\empty,
        axis on top,
        legend style={at={(0.5,1.02)}, anchor=south, legend columns=2},
        ytick={10, 100, 1000, 10000, 100000, 1000000},
        extra x ticks={-1.5,0,1.5},
        extra x tick labels={\(\frac{p_{\mathrm{dex}}}{p_{\mathrm{cex}}} < 1\), \(1\),\(\frac{p_{\mathrm{dex}}}{p_{\mathrm{cex}}} > 1\)},
        extra x tick style={
            major tick length=0pt,
            tick label style={yshift=-8pt},
        }
    ]
    \addlegendimage{area legend, fill=blue, draw=none, opacity=0.15}
    \addlegendentry{No-arbitrage zone}
    
    \addlegendimage{area legend, fill=green, draw=none, opacity=0.15}
    \addlegendentry{Joint profit zone}

    \addlegendimage{area legend, fill=red, draw=none, opacity=0.15}
    \addlegendentry{IL zone}
        
    \pgfmathsetmacro{\YMAX}{10000000}
    
    \path[name path=left_top]    (axis cs:-3,\YMAX) -- (axis cs:-1.5,\YMAX);
    \path[name path=left_bottom] (axis cs:-3,0)     -- (axis cs:-1.5,0);
    \addplot[red!15, draw=none, forget plot] fill between[of=left_top and left_bottom];
    
    \path[name path=mid_top]    (axis cs:-1.5,\YMAX) -- (axis cs:-0.5,\YMAX);
    \path[name path=mid_bottom] (axis cs:-1.5,0)     -- (axis cs:-0.5,0);
    \addplot[green!15, draw=none, forget plot] fill between[of=mid_top and mid_bottom];

    \path[name path=mid_top]    (axis cs:0.5,\YMAX) -- (axis cs:1.5,\YMAX);
    \path[name path=mid_bottom] (axis cs:0.5,0)     -- (axis cs:1.5,0);
    \addplot[green!15, draw=none, forget plot] fill between[of=mid_top and mid_bottom];

    \path[name path=mid_top]  (axis cs:-0.5,\YMAX) -- (axis cs:0.5,\YMAX);
    \path[name path=mid_bottom] (axis cs:-0.5,0)     -- (axis cs:0.5,0);
    \addplot[blue!15, draw=none, forget plot] fill between[of=mid_top and mid_bottom];
    
    \path[name path=right_top]    (axis cs:1.5,\YMAX) -- (axis cs:3,\YMAX);
    \path[name path=right_bottom] (axis cs:1.5,0)     -- (axis cs:3,0);
    \addplot[red!15, draw=none, forget plot] fill between[of=right_top and right_bottom];
    
    \addplot+[ybar, bar width=25pt, fill=blue!40] coordinates {
        (-2, 37)
        (-1, 133)
        (0, 1063594)
        ( 1, 101)
        ( 2, 36)
    };
    
    \draw[black, very thick] (axis cs:0,0) -- (axis cs:0,\YMAX);
    
    \end{axis}
    \end{tikzpicture}
    \caption{Transaction distribution across profitability zones for Pool A (0.03\% fee). The central blue region shows no-arbitrage blocks, green regions represent win-win trades, and red regions indicate IL-causing arbitrage.}
    \label{fig:ig_distr}
\end{figure}
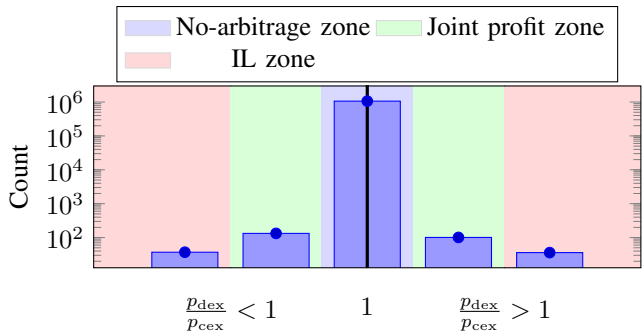

Figure~\ref{fig:ig_distr} shows the distribution of transactions across profitability zones for Pool A~(0.03\% fee). Consistent with our theoretical predictions, 76\% of all transactions occurred within the joint profitability~(IG) zone where both the arbitrageur and LPs profit and 24\% in IL zone, but there were no transactions in most of the blocks, which corresponds to the no-arbitrage zone. This distribution validates our model's key insight: properly calibrated fees can create a sustainable win-win regime even under competitive market conditions.

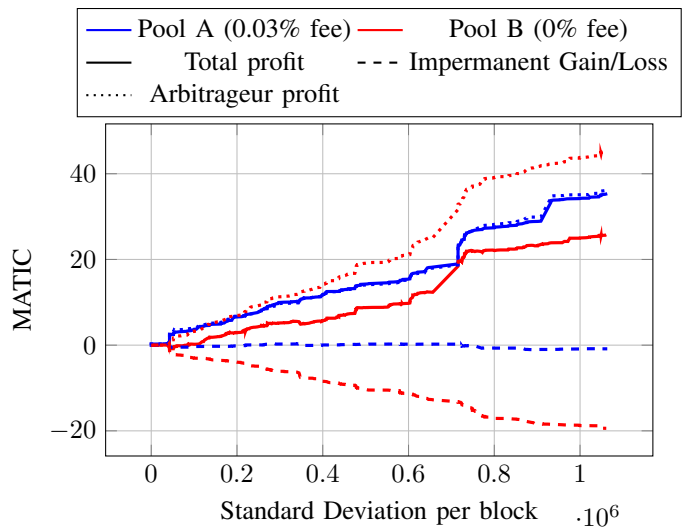
\begin{figure}[t]
\centering
\begin{tikzpicture}
\begin{axis}[
  width=\linewidth,
  height=6cm,
  grid=both,
  xlabel={Standard Deviation per block},
  ylabel={MATIC},
legend style={at={(0.5,1.02)}, anchor=south, legend columns=2},
  cycle list={}
]

\addlegendimage{thick, solid, blue}
\addlegendentry{Pool A (0.03\% fee)}

\addlegendimage{thick, solid, red}
\addlegendentry{Pool B (0\% fee)}

\addlegendimage{thick, solid, black}
\addlegendentry{Total profit}

\addlegendimage{thick, dashed, black}
\addlegendentry{Impermanent Gain/Loss}

\addlegendimage{thick, dotted, black}
\addlegendentry{Arbitrageur profit}

\addplot+[no marks, very thick, solid, blue] 
table[col sep=comma, x=block_no, y=total_profit]{data/fee_pool_profit.csv};

\addplot+[no marks, very thick, solid, red] 
table[col sep=comma, x=block_no, y=total_profit]{data/zero_pool_profit.csv};

\addplot+[no marks, very thick, dashed, blue] 
table[col sep=comma, x=block_no, y=IL]{data/fee_pool_profit.csv};

\addplot+[no marks, very thick, dashed, red] 
table[col sep=comma, x=block_no, y=IL]{data/zero_pool_profit.csv};

\addplot+[no marks, very thick, dotted, blue] 
table[col sep=comma, x=block_no, y=arb_profit]{data/fee_pool_profit.csv};

\addplot+[no marks, very thick, dotted, red] 
table[col sep=comma, x=block_no, y=arb_profit]{data/zero_pool_profit.csv};

\end{axis}
\end{tikzpicture}
\caption{Profit evolution over the 26-day experiment. Pool A (with 0.03\% fee) shows impermanent gain (IG) fluctuating around 0 MATIC with a final result of -1 MATIC, and arbitrage profit reaching up to 36 MATIC. Pool B (zero fee) exhibits impermanent loss (negative IG) up to -19 MATIC and arbitrage profit up to 45 MATIC. The combined profit (arbitrage + IG) is positive in both cases, with Pool A achieving approximately 35 MATIC and Pool B 26 MATIC.}
\label{fig:profit_evolution}
\end{figure}

The temporal evolution of profits provides further insight into the system dynamics. Figure~\ref{fig:profit_evolution} tracks the key metrics over the experiment duration. Both pools exhibited stable results. Pool A (with 0.03\% fee) shows impermanent gain (IG) remaining close to zero, concluding at -1 MATIC, while its arbitrage profit reaches up to 36 MATIC. Pool B (zero fee) demonstrates impermanent loss (negative IG) up to -19 MATIC, with arbitrage profit peaking at 45 MATIC. The resulting sum of arbitrage and IG is positive in both cases: Pool A achieves a combined profit of approximately 35 MATIC, while Pool B reaches about 26 MATIC. This demonstrates that a fair distribution of profit between arbitrageur and LP, as facilitated by a positive fee structure in Pool A, can result in simultaneous stable profit for both parties, underscoring that collaboration matters. Furthermore, the positive fee in Pool A acts as a regularization mechanism, preventing low-profit arbitrage transactions that would be risky due to price volatility.

We note an important observation: the zero-fee pool generated 50\% more arbitrage transactions and 25\% higher arbitrageur profit than the fee-charging pool. This demonstrates that eliminating fees increases arbitrage frequency and potential arbitrageur profits. In principle, an arbitrageur operating in a zero-fee pool could share a portion of their increased profits with LPs, potentially creating a win-win scenario even without fees. However, such profit-sharing arrangements require explicit coordination or protocol mechanisms that were not implemented in our experiment. In our baseline zero-fee setup without profit sharing, LPs bore the full brunt of impermanent loss while arbitrageurs captured all gains.

The comparative performance metrics reveal the economic implications of our framework. Pool A achieved a combined annualized return of 3.2\% APR, outperforming Pool B's 2.4\% APR by 33\%. However, both private pools underperformed relative to a comparable public pool (contract address \seqsplit{0x604229c960e5cacf2aaeac8be68ac07ba9df81c3}) which yielded a 9.5\% APR. This performance gap highlights the competitive efficiency of open, multi-arbitrageur markets in extracting value from price discrepancies. Nevertheless, the superior performance of Pool A relative to Pool B stems from the symbiotic relationship enabled by the fee structure: LPs provide compensation (through fees) to arbitrageurs for stabilizing the pool, while arbitrageurs return value to LPs by maintaining prices within the IG zone. The zero-fee pool generated 1.5$\times$ more arbitrage transactions, confirming that the absence of fees incentivizes greater arbitrage activity. This suggests that explicit, protocol-enforced profit-sharing mechanisms in zero-fee pools could potentially bridge the performance gap with public pools by redistributing a portion of the higher arbitrageur profits to LPs.

These experimental results validate several key aspects of our theoretical framework:
\begin{enumerate}
    \item \textbf{Existence of IG zones:} The concentration of transactions in the win-win region for Pool A confirms that properly calibrated fees create sustainable profitability zones where both LPs and arbitrageurs benefit simultaneously.
    
    \item \textbf{Risk reduction:} The zero-fee pool (Pool B) exhibited impermanent loss up to -20 MATIC, while the fee-charging pool (Pool A) maintained IG near zero, demonstrating that fees provide a stabilizing regularization effect against volatile, low-profit arbitrage.
    
    \item \textbf{Collaborative value creation:} The combined profit was positive in both pools, with Pool A achieving a higher total ($\approx$35 MATIC vs. $\approx$25 MATIC). This indicates that the fee structure in Pool A fostered a more efficient collaboration, leading to greater overall value extraction from price discrepancies.
    
    \item \textbf{Arbitrage volume vs. fee trade-off:} Zero-fee arbitrage generated greater absolute profit for the arbitrageur (+45~MATIC vs. +36 MATIC), confirming that the absence of fees incentivizes more arbitrage activity. However, this does not automatically translate to better LP outcomes without explicit coordination.

    \item \textbf{Practical feasibility:} The successful implementation proves that a collaborative LP-arbitrageur system is practically achievable. The results suggest that protocol designs incorporating fee-based regularization can create more sustainable and predictable profit zones for LPs.
\end{enumerate}

The experimental results demonstrate two distinct pathways to sustainable liquidity provision: (1) Fee-based coordination (Pool A) creates an automatic win-win zone through built-in fee mechanisms, providing predictable returns with lower volatility. (2) Zero-fee operation (Pool B) generates more arbitrage opportunities and higher potential arbitrageur profits, but requires explicit profit-sharing arrangements to benefit LPs. Our theoretical framework works well for fee-based pools despite modeling simplifications, while the zero-fee case highlights the need for additional coordination mechanisms to realize its full potential.

It is important to acknowledge that our theoretical model makes several simplifying assumptions that differ from our experimental setup. The theoretical analysis assumes an ideal arbitrageur operating between a DEX and a CEX with perfect monitoring and immediate execution every block. In contrast, our experimental implementation used a real arbitrage bot operating between DEX pools (our private pool and public DEXs) with practical constraints including network latency, gas price competition, and occasional missed opportunities. Furthermore, the exclusive whitelist access eliminated competition from other arbitrageurs, creating conditions different from open markets. Despite these differences in assumptions and implementation, the experimental results align closely with the theoretical predictions--particularly the existence of the IG zone and the win-win transactions comprising 76\% of all trades. This demonstrates that while the theoretical model simplifies reality, it captures the essential dynamics and provides a useful approximation for real-world scenarios.

The experiment also revealed an important practical consideration: the 0.03\% fee, while sufficient to create an IG zone, resulted in relatively infrequent arbitrage activity compared to the zero-fee pool. This suggests a trade-off between fee income frequency and protection against IL--a consideration that could inform dynamic fee strategies in future implementations.

A key limitation of our experimental setup is the use of a single, whitelisted arbitrageur, which eliminates competition and MEV considerations present in public pools. While this simplifies analysis, future work should explore multi-arbitrageur competitive equilibria and their impact on the IG zone boundaries. Additionally, our 26-day experiment captures limited market regimes; longer-term studies across varying volatility environments would strengthen the empirical validation.

\section{Conclusion}\label{sec:conclusion}

This paper presents a comprehensive framework for quantifying profitability zones in Automated Market Maker-based decentralized exchanges, transforming the conventional view of impermanent loss from an unavoidable risk to a manageable parameter. We bridge theoretical modeling with empirical validation to demonstrate that properly calibrated fee structures can create sustainable win-win scenarios for both liquidity providers and arbitrageurs.

Our primary contribution is the formal characterization of Impermanent Gain zones--price ratio intervals where arbitrage transactions benefit both parties simultaneously. For constant product AMMs like Uniswap V2 and Balancer, we derived explicit analytical boundaries for these zones and developed a probabilistic risk framework that translates these boundaries into practical metrics for liquidity providers. By modeling external price dynamics as Geometric Brownian Motion, we enable LPs to quantify impermanent loss probability and duration, providing a principled basis for fee optimization based on market volatility and risk tolerance.

Through controlled on-chain experiments with private pools, we empirically validated our theoretical framework. The experimental results confirmed that 76\% of arbitrage transactions occurred within the predicted win-win zone when using a 0.03\% fee, demonstrating that collaborative LP-arbitrageur systems can eliminate impermanent loss while enhancing combined returns. Crucially, our experiment revealed a fundamental trade-off: zero-fee pools generate 50\% more arbitrage transactions and 25\% higher arbitrageur profits (45 MATIC vs. 36 MATIC), yet the combined total wealth (arbitrageur profit + LP impermanent gain) is 40\% higher in the fee-charging pool (35 MATIC vs. 25 MATIC). This demonstrates that fee-based regularization enhances overall economic efficiency and common wealth creation, not merely redistributing value between participants. While zero-fee pools create greater arbitrageur value extraction potential, this comes at the expense of total system efficiency, highlighting that fees serve as coordination mechanisms that optimize collective outcomes.

The implications of this work extend to multiple stakeholders in decentralized finance. Protocol designers gain a quantitative framework for optimizing fee structures based on measurable market risk, potentially reducing toxic order flow and improving overall market stability. Liquidity providers acquire tools for assessing pool risk and selecting positions aligned with their risk-return preferences, moving beyond heuristic approaches to liquidity provision. Our analysis of private versus public pool performance provides guidance for sophisticated LPs considering exclusive arrangements with arbitrageurs, particularly in emerging markets or for exotic trading pairs where these advantages are most pronounced.

Several promising directions emerge for future research. Extending this framework to concentrated liquidity AMMs like Uniswap V3 would address the more complex zonal boundaries created by liquidity fragmentation. Investigating multi-asset pools and cross-pool arbitrage could reveal additional dimensions of the LP-arbitrageur relationship. Developing dynamic fee mechanisms that adjust based on real-time volatility estimates and arbitrage competition could enhance protocol responsiveness to changing market conditions. Finally, exploring decentralized profit-sharing mechanisms for zero-fee pools--such as bonding curves for arbitrageur-LP agreements or protocol-level profit distribution systems--could enable zero-fee pools to realize their full potential without relying on trusted relationships, bringing the benefits of private pool arrangements to public settings. 

By bridging rigorous theoretical analysis with practical on-chain experimentation, this work advances decentralized finance toward more sustainable and predictable liquidity provision. Rather than viewing impermanent loss as an inevitable hazard, our framework treats it as a quantifiable parameter within a win-win market design. Regulatory considerations will also shape the evolution of DEX design and adoption, as cryptocurrency regulation impacts trading markets and compliance requirements~\cite{Feinstein2021}. As DeFi continues to mature, such quantitative approaches will be essential for building robust, efficient, and equitable financial systems on blockchain infrastructure.

\bibliographystyle{IEEEtran}
\bibliography{references}
\end{document}